\documentclass[
  reprint,
  onecolumn,
  pre,
  aps,
  amsmath,
  amssymb,
  superscriptaddress,
  floatfix
]{revtex4-2}
\usepackage{amssymb,amsmath,amsfonts,multirow,rotate,color,graphicx}
\usepackage{hyperref}
\usepackage[page]{appendix}
\usepackage{tikz,lipsum,lmodern}
\usepackage[most]{tcolorbox}

\usepackage{graphicx}

\usepackage{lineno}

\usepackage{times}

\usepackage{booktabs}

\newcounter{lastnote}

\renewcommand{\[}{\left[}

\newcommand{\bra}[1]{\left\langle #1 \right|}
\newcommand{\ket}[1]{\left| #1 \right\rangle}
\newcommand{\braket}[2]{\left\langle #1 \,\middle|\, #2 \right\rangle}

\usepackage{xcolor}
\definecolor{RoyalBlue}{HTML}{4169e1}
\definecolor{ForestGreen}{HTML}{228b22}

\usepackage{setspace}
\usepackage{subfigure}
\usepackage{comment}

\begin{document}

\title{Information-thermodynamic principles unveil the response of complex networks to perturbations}

\author{Andrea Filippo Beretta}
\email{andreafilippo.beretta@studenti.unipd.it}
\affiliation{Department of Physics and Astronomy “Galileo Galilei”, University of Padua, Via F. Marzolo 8, 35131 Padova, Italy}

\author{Manlio De Domenico}
\email{manlio.dedomenico@unipd.it}
\affiliation{Department of Physics and Astronomy “Galileo Galilei”, University of Padua, Via F. Marzolo 8, 35131 Padova, Italy}
\affiliation{Padua Center for Network Medicine, University of Padua, 35131 Padova, Italy}
\affiliation{Padova Neuroscience Center, University of Padua, Via G. Orus 2, 35131 Padova, Italy}
\affiliation{Istituto Nazionale di Fisica Nucleare, Sez. Padova, Padua, Italy}

\date{\today}

\baselineskip24pt

\begin{abstract}

Understanding how complex networks respond to perturbations is key to uncovering their functional organization. We show that, for a maximum-entropy perturbation prior, standard diffusion cannot increase response entropy, a measure of the diversity of the system’s response. To overcome this limitation, we develop a density-matrix-based information thermodynamics of perturbation responses along arbitrary dynamical trajectories. We show that this framework is grounded in an information-theoretic principle of least dissipation and introduces a generalized free energy that quantifies deviations from an instantaneous diffusive reference. We then illustrate the formalism using nonlinear complex networks, revealing regimes of enhanced response diversity that cannot be captured by standard diffusive descriptions.

\end{abstract}

\maketitle

\section{Introduction} 

A defining feature of complex systems is not only their structure, but also their response to perturbations \cite{kauffman2004ensemble,de2005spectral,maslov2007spreading,hens2019spatiotemporal}. Systems with similar unperturbed trajectories may react very differently to small disturbances, revealing latent structural and dynamical properties \cite{gao2016universal,dobson2007complex}. Characterizing perturbation responses therefore provides a fundamental lens on the organization and function of complex networks, since perturbation propagation reveals how information, uncertainty, and functional variability are redistributed by the interplay between structure and dynamics~\cite{busiello2017explorability,ghavasieh2024diversity,gao2016universal,hens2019spatiotemporal}. 

Standard diffusive dynamics offers a tractable class of network processes: its evolution is linear and governed by a time-independent, often symmetric, operator, under which perturbations evolve as well. Diffusion models phenomena ranging from consensus formation \cite{delellis2010analysis} to epidemic spreading \cite{fisher1937wave,brockmann2013hidden} and migration \cite{hastings1993complex}, and often approximates nonlinear dynamics near steady states \cite{van1992stochastic}. Recently, a density matrix formalism \cite{de2016spectral}, analogous to equilibrium statistical mechanics, enabled a thermodynamic description of networks coupled to diffusion. This diffusive information--thermodynamic framework has been applied to network classification~\cite{ghavasieh2021multiscale,su2021identification}, phase transitions~\cite{villegas2022laplacian}, centrality~\cite{ghavasieh2021unraveling}, robustness~\cite{ghavasieh2023dismantling}, functional fragility~\cite{ghavasieh2023generalized}, structural sparsity~\cite{ghavasieh2024diversity}, and renormalization group approaches~\cite{villegas2023laplacian,nurisso2025higher}. 

Despite these advances, it remains unclear whether diffusion provides an exhaustive description, or whether a more general thermodynamic framework is needed to capture behaviors beyond it. This is especially relevant for real networks, which often operate in dynamical regimes far from steady states. As non-equilibrium thermodynamics and statistical mechanics \cite{j2007statistical,onsager1931reciprocal,onsager1931reciprocal2} describe phenomena inaccessible to equilibrium approaches, an analogous extension may be required for network responses governed by general time-dependent and non-normal operators \cite{asllani2018structure}.

In this article, we explore the theoretical limits of the diffusive approach commonly adopted in the network density matrix literature, which relies on a time-independent normal evolution operator and on maximum-entropy assumptions for the prior distribution of perturbations. In particular, we show that, under these common assumptions, the entropy derivative is constrained to be non-positive. This contrasts with real systems, which may maintain or increase functional diversity. We therefore construct a general information-thermodynamic framework capable of describing the response to perturbations at any point along a complex system trajectory and with a time-dependent non-normal evolution operator. This framework is based on an information-theoretic least dissipation principle, analogous in spirit to the Onsager–Rayleigh principle, and establishes a formal correspondence between non-equilibrium thermodynamics and network response theory.

\section{Results}

\subsection{Network density matrices} 

The network density matrix formalism~\cite{de2016spectral,ghavasieh2023generalized} provides an information--thermodynamic framework to describe perturbation propagation in an $N$-dimensional complex network with adjacency matrix $\mathbf{A}$, whose state $\ket{\mathbf{x}(t)}$ follows the non-linear dynamics $\frac{d}{dt}\ket{\mathbf{x}(t)}=\mathbf{v}(\ket{\mathbf{x}(t)};\mathbf{A})$. Perturbations are small shifts $\ket{\delta\mathbf{x}(t)}$ of the system state $\ket{\mathbf{x}(t)}$ (see Appendix~\ref{app::response}).
\begin{table*}
\caption{\label{tab:dynamics}
Dynamical equations of complex networks considered in this article.}
\begin{ruledtabular}
\begin{tabular}{lll}
Symbol & Dynamics & $\dot{x}_i =$ \\
\hline

$\mathbb{E}$ & Epidemics (SIS) &
$-B x_i + R \sum_j A_{ij}\,(1-x_i)x_j$ \\[1.2ex]


$\mathbb{NV}$ & Noisy voter &
$A - B x_i + \dfrac{C}{k_i}\sum_j A_{ij} x_j$ \\[1.2ex]


$\mathbb{P}$ & Population &
$-B x_i^{\,b} + R \sum_j A_{ij} x_j^{\,a}$ \\[1.2ex]

$\mathbb{N}$ & Neuronal &
$-x_i + R \sum_{j=1}^{N} A_{ij}\,\tanh(x_j)$ \\[1.2ex]

$\mathbb{B}$ & Biochemical &
$F - B x_i - R \sum_j A_{ij} x_i x_j$ \\
\end{tabular}
\end{ruledtabular}
\end{table*}
Denoting by $\ket{i}$ the canonical basis vector associated with node $i$, the response of individual nodes to a perturbation mode $\ket{\delta\mathbf{x}(t)}$ is quantified through its projection onto this basis. The absolute response intensity is
$|\delta x_i (t)|=\sqrt{|\braket{\delta\mathbf{x}(t)}{i}|^2},$
while the dynamical correlation between nodes $j$ and $i$ is
$\delta x_j (t) \delta x_i (t)=\langle j |\delta\mathbf{x}(t) \rangle \langle \delta\mathbf{x}(t)|i \rangle$.
Both are encoded in the local propagator $\mathbf{u}(t) = \ket{\delta\mathbf{x}(t)} \bra{\delta\mathbf{x}(t)}$,
which describes the spreading of a single perturbation mode.

A single perturbation generally cannot characterize the full network response, since different initial directions may produce distinct dynamics and the actual perturbation is often unknown. We therefore use an ensemble of $M$ possible initial perturbations,
$\Psi=\{\delta \mathbf{x}^\alpha (t_0)\}_{\alpha=0,...,M-1}$,
where each mode $\delta \mathbf{x}^\alpha$ occurs at time $t_0$ with probability $p_\alpha$ and evolves under the Jacobian $\mathbf{J}=\frac{\partial\mathbf{v}}{\partial \mathbf{x}}$ (see Tab.~\ref{tab:dynamics}). To span all perturbation directions, we choose $M=N$ orthogonal modes.

The propagation of information through the network is then described by the statistical propagator $\boldsymbol{\mathcal{U}}(t) = \sum_{\alpha=0}^{N-1} p_\alpha \mathbf{u}^\alpha(t)$, where $\mathbf{u}^\alpha(t)$ is the local propagator of mode $\alpha$. Its trace,
$Z(t)\equiv\operatorname{Tr}(\boldsymbol{\mathcal{U}}(t)),$
equals the total weighted ensemble strength. The normalization of $\boldsymbol{\mathcal{U}}(t)$ gives the network density matrix
\begin{equation}
\boldsymbol{\rho}(t)=\frac{\boldsymbol{\mathcal{U}}(t)}{Z(t)}.
\end{equation}
This positive, unit-trace operator encodes the distribution of the perturbation ensemble across network nodes and modes.

Without prior information on $p_\alpha$, the maximum entropy principle~\cite{jaynes1957information,jaynes1957information2} assigns probability $1/N$ to each mode. A standard choice for $\Psi$ is the canonical basis, with perturbations initially localized on single nodes~\cite{de2016spectral,villegas2022laplacian,ghavasieh2023generalized,villegas2023laplacian,ghavasieh2024diversity}. To isolate the effect of the evolution operator, throughout the main text we adopt this maximum-entropy prior for the initial perturbation ensemble, $\boldsymbol{\rho}(t_0)=\mathbf{I}/N$, unless otherwise specified. This keeps the initial ensemble fixed when comparing different evolution operators and avoids introducing an additional directional bias through the perturbation prior. The role of non-uniform priors, including their possible effect on the sign of the entropy rate, is discussed in Appendix~\ref{app::prior}.

For diffusive dynamics, this framework is formally connected to equilibrium statistical mechanics and the Gibbs ensemble of densities~\cite{ghavasieh2022statistical} (see Appendix~\ref{app::correspondence}). Moreover, by selecting this uniform prior for the initial perturbation distribution, the density matrix formally coincides with the normalized left Cauchy--Green tensor of classical dynamical systems theory~\cite{politi2013lyapunov}. Therefore, the density-matrix representation does not introduce additional microscopic information with respect to tangent-space dynamics, but changes the descriptive level by recasting the collective response to perturbations as a statistical state. This representation makes it possible to summarize high-dimensional response dynamics through a small set of macroscopic thermodynamic observables and to apply statistical and variational tools to characterize the joint effects of network structure and dynamics across scales.

\subsection{The density matrix flow equation and entropy evolution}

The first step toward a response theory for complex interconnected systems along general trajectories is to understand how the evolution deforms the density operator $\boldsymbol{\rho}(t)$ describing response modes. For perturbations evolving under a time-dependent, generally non-symmetric $\mathbf{J}(\mathbf{x}(t))$, differentiating the density matrix (Appendix~\ref{app::density}) yields the equation governing its flow on the manifold of all admissible density matrices:
\begin{equation} \label{Liouville}
\dot{\boldsymbol{\rho}}=\mathbf{J} \boldsymbol{\rho}+ \boldsymbol{\rho} \mathbf{J}^T-2 \boldsymbol{\rho} \operatorname{Tr}(\boldsymbol{\rho}\mathbf{J}).
\end{equation}
The first two terms describe the local deformation of $\boldsymbol{\rho}$ induced by $\mathbf{J}(\mathbf{x}(t))$, including stretching and directional changes of perturbation modes. The third term preserves $\operatorname{Tr}(\boldsymbol{\rho})=1$ and thus contains $\operatorname{Tr}(\boldsymbol{\rho}\mathbf{J})$, linked to the local stretching rate of an infinitesimal volume around the trajectory.

We then exploit the flow equation to explore the role of symmetric and antisymmetric dynamical features \cite{bottcher2024complex}. Decomposing $\mathbf{J}$ into its symmetric part $\mathbf{D}$ and antisymmetric part $\mathbf{C}$ gives
\begin{equation}
\dot{\boldsymbol{\rho}}=\{\mathbf{D},\boldsymbol{\rho}\}-2 \boldsymbol{\rho} \operatorname{Tr}(\boldsymbol{\rho}\mathbf{D})+[\mathbf{C},\boldsymbol{\rho}],
\end{equation}
where $\{\cdot,\cdot\}$ and $[\cdot,\cdot]$ denote anticommutator and commutator. Eigenvalue and eigenvector perturbation theory \cite{li2006matrix} (Appendix~\ref{app::density}) shows that $\mathbf{D}$ affects both eigenvalues and eigenvectors of $\boldsymbol{\rho}$, while $\mathbf{C}$ only rotates eigenvectors unitarily. Thus, $\mathbf{D}$ represents the dissipative part of the evolution, changing the relative relevance of perturbations, whereas $\mathbf{C}$ is conservative. Consistently, only $\mathbf{D}$ enters the normalization term, since $\operatorname{Tr}(\boldsymbol{\rho}\mathbf{J})=\operatorname{Tr}(\boldsymbol{\rho}\mathbf{D})$. The diffusive case is recovered by setting $\mathbf{C}=0$ and assuming time independence of $\mathbf{D}$.

The conservative part can always be removed by rotating $\boldsymbol{\rho}(t)$ and $\mathbf{D}(t)$ through $\mathbf{O}(t,t_0)= \mathcal{T} { \exp[\int_{t_0}^{t}ds\, \mathbf{C}(s)]}$, where $\mathcal{T}$ is the time-ordered exponential. In this dissipative frame, $\boldsymbol{\sigma}(t)\equiv \mathbf{O}^{-1}\boldsymbol{\rho}(t)\mathbf{O}$ evolves only under $\mathbf{Q}(t)\equiv \mathbf{O}^{-1}\mathbf{D}(t)\mathbf{O}$:
\begin{equation}
\dot{\boldsymbol{\sigma}}=\{\mathbf{Q},\boldsymbol{\sigma}\}-2 \boldsymbol{\sigma} \operatorname{Tr}(\boldsymbol{\sigma}\mathbf{Q}).
\end{equation}
This rotation preserves relevant thermodynamic quantities such as the Von Neumann entropy $S$ and $\operatorname{Tr}(\boldsymbol{\rho}\mathbf{J})$, since they are invariant under unitary transformations. Despite this formal mapping, the time dependence and the non-normality of the evolution operator in the general case mark a fundamental difference from the time-independent and symmetric diffusive case.

We exploit the flow equation to study the evolution of the Von Neumann entropy $S= -\operatorname{Tr}(\boldsymbol{\rho}\log \boldsymbol{\rho})$, which measures the functional diversity of the response through the relative importance of its modes. Our goal is to identify differences from the time-independent diffusive case. Substituting the flow equation into the entropy derivative yields
\begin{equation}\label{ent_dev}
\dot{S}= -2 \,\operatorname{Cov}_{\boldsymbol{\rho}}(\mathbf{D},\log\boldsymbol{\rho}),
\end{equation}
where $\operatorname{Cov_{\boldsymbol{\rho}}(\cdot,\cdot)}$ denotes the covariance with respect to $\boldsymbol{\rho}$ \cite{gibilisco2008quantum}. By the Cauchy--Schwarz inequality, we then obtain
\begin{equation} \label{bound}
|\dot{S}|\leq 2\sqrt{\operatorname{Var}_{\boldsymbol{\rho}}(\mathbf{D})
\operatorname{Var}_{\boldsymbol{\rho}}(\log \boldsymbol{\rho})}.
\end{equation}
This is an upper bound on the entropy rate \cite{nicholson2020time} and sets a scale-dependent limit to changes in functional diversity. Given a scale $t$, we define the instantaneous Gibbsian reference as $\boldsymbol{\gamma}(t)= e^{2t \mathbf{D}(t)}/\operatorname{Tr}(e^{2t \mathbf{D}(t)})$, obtained by freezing the symmetric operator $\mathbf{D}(t)$ at that time. We call $\boldsymbol{\gamma}(t)$ the instantaneous diffusive reference, interpreting it as the effective diffusive process locally approximating the real dynamics. The bound in Eq.~(\ref{bound}) is saturated when $\log\boldsymbol{\rho}=a\mathbf{D}+b\mathbf{I}$, with $a,b\in\mathbb{R}$. As a special case, this occurs when the actual density operator coincides with $\boldsymbol{\gamma}(t)$ at the corresponding scale.

\begin{figure*}
\includegraphics[width=\textwidth]{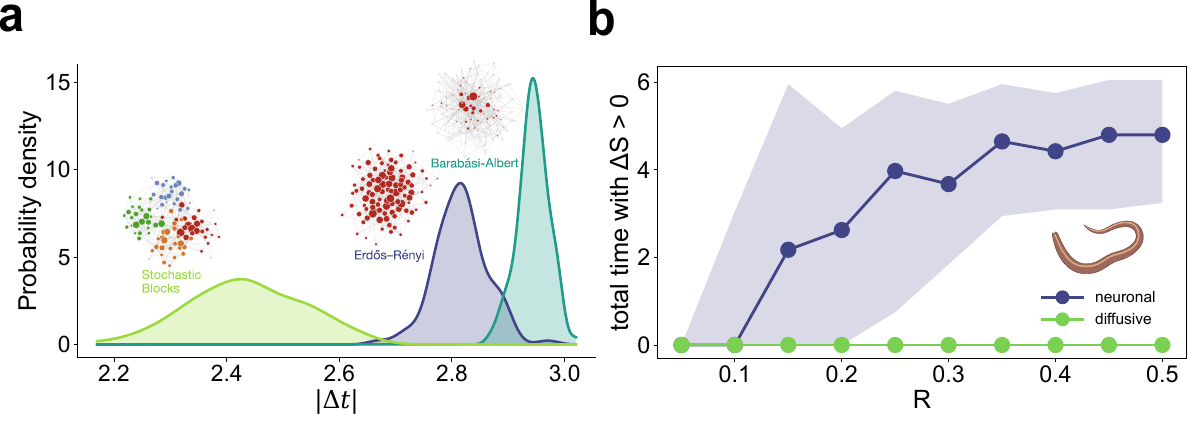}
\caption{\label{fig:entropy_rate}
\textbf{Beyond diffusion: nonlinear dynamics systematically alters the evolution of the entropy rate.} \textbf{(a)} The same network can display different entropy production rate profiles under different dynamics. The figure shows the temporal distance $|\Delta t|$ between the peaks of $|\dot{S}|$ for Laplacian diffusion and biochemical dynamics ($\mathbb{B}$) in Erdős--Rényi, Barabási--Albert, and stochastic block model networks with average degree $\langle k \rangle \sim 8$ and $N=100$, over $100$ realizations each. All the networks exhibit distributions not centered at zero, indicating a clear difference between diffusive and non-diffusive dynamics. 
\textbf{(b)} The \textit{C. elegans} nervous system network \cite{white1986structure} is used to compare the evolution of normalized entropy under Laplacian diffusion and neuronal dynamics $(\mathbb{N})$ across different coupling strengths $R$, using a low-activity initial condition. For neuronal dynamics, $30$ inhibitory-link assignments are considered per $R$, with inhibitory probability $26\%$ \cite{ghavasieh2023generalized}. The range of values of $R$ explored spans the dynamical scale around $1/\langle \operatorname{Re}(\lambda_{max})\rangle$, where $\langle \operatorname{Re}(\lambda_{\max}) \rangle \sim 9.7$ denotes the ensemble-averaged real part of the largest eigenvalue of the signed adjacency matrix. This choice is motivated by the widespread interest in critical and near-critical dynamical regimes in the study of neuronal dynamics \cite{hesse2014self}. Unlike diffusion, nonlinear neuronal dynamics can show regions with $\dot{S}>0$; the figure reports the median total time with positive entropy derivative across coupling strengths, with shading between the $16\%$ and $84\%$ quantiles. These regions are consistently present at higher $R$. }
\end{figure*}

Aside from the comparison with its instantaneous diffusive counterpart, we also contrast the entropy derivative of a system evolving under non-linear dynamics with that of a network with the same adjacency matrix evolving under Laplacian diffusion (Fig.~\ref{fig:entropy_rate}--\textbf{a}).

\textit{Network H-theorem}\textemdash Both diffusive approximations of complex network dynamics discussed in the previous section exhibit deviations from their non-diffusive counterparts. Although these differences can be significant, they do not seem to involve features that are, in principle, theoretically incompatible with a time-independent and symmetric diffusive approximation.

However, one phenomenon cannot be captured in such a framework: entropy increase, i.e. a positive entropy derivative $\dot{S}>0$. In real systems, we observe phases where the diversity of responses to perturbations increases, due to both structural and dynamical effects. In the time-independent and symmetric diffusive case, instead, the entropy derivative is always non-positive. Indeed, for a constant and symmetric evolution operator $\mathbf{D}$ and a maximum--entropy prior, Eq.~(\ref{ent_dev}) reduces to $\dot{S}= -4 \,t \operatorname{Var}_{\boldsymbol{\rho}}(\mathbf{D})$; since the variance is non-negative, $\dot{S}\leq0$ (Appendix~\ref{app::network}). This is analogous to the Boltzmann H-theorem \cite{boltzmann1970weitere}, but with an inverted sign due to the different interpretation of entropy in our context, where we start from a maximum--entropy state. This bound has a clear physical interpretation: if the evolution operator is constant and symmetric and we assume a maximum--entropy prior, the diversity of the response to perturbations of a complex network cannot increase in time. For this reason, a non-diffusive information-thermodynamic framework is required to fully explain the response to perturbations along a general dynamical trajectory, where a non-uniform perturbation prior, time dependence, or non-normality of the evolution operator may each yield $\dot{S}>0$. Indeed, Eq.~(\ref{ent_dev}) shows that the sign of the entropy rate is not determined by the instantaneous symmetric operator $\mathbf{D}$ alone, but by its relation to the current density operator, which contains the accumulated effect of both the initial perturbation ensemble and the preceding dynamics. Let
$\boldsymbol{\rho}=\sum_i p_i |p_i\rangle\langle p_i|$ and define
$d_i^{(\rho)}\equiv\langle p_i|\mathbf{D}|p_i\rangle$ and $\left\langle\mathbf{D}\right\rangle= \operatorname{Tr}(\mathbf{D} \boldsymbol{\rho})$, where
$d_i^{(\rho)}$ is the instantaneous stretching rate along the $i$th response direction. Then, equation~(\ref{ent_dev}) can be rewritten as
\begin{equation}
    \dot{S}=-2\sum_i p_i \left(\log p_i-\left\langle\log\boldsymbol{\rho}\right\rangle\right)\left(d_i^{(\boldsymbol{\rho})}-\left\langle\mathbf{D}\right\rangle\right).
\end{equation}
Hence, the sign of $\dot{S}$ is controlled by the correlation between the accumulated relevance of each response mode, encoded by $\log p_i$ and shaped by both the initial prior and the previous evolution along the trajectory, and its instantaneous stretching, represented by $d_i^{(\boldsymbol{\rho})}$. In particular, $\dot{S}>0$ when these two quantities are negatively correlated. Physically, this occurs when modes that were previously weakly represented undergo comparatively stronger instantaneous stretching, thereby redistributing the response toward previously suppressed directions. This mechanism is absent in the standard diffusive case, for which
$\log\boldsymbol{\rho}=2t\mathbf{D}-(\log Z)\,\mathbf{I}$, so that mode relevance remains locked to the fixed stretching rates and cannot become anticorrelated with them, whereas such an anticorrelation may occur whenever this relation no longer holds (Appendices~\ref{app::prior}--~\ref{app::time}--~\ref{app::normality}).

To test whether this thermodynamic signature can be associated with concrete dynamical features, we consider the \textit{C.~elegans} neuronal network with nonlinear neuronal dynamics $(\mathbb{N})$. In the low-activity regime, the system has a linear-stability threshold around $R_c\simeq 0.1$, above which the Jacobian can acquire eigenvalues with positive real part, signaling the onset of locally unstable modes in the linearized dynamics. We therefore examine whether the occurrence of $\dot{S}>0$ changes across this threshold. In this system, the total time spent with $\dot{S}>0$ becomes appreciable above $R_c$ and increases at stronger coupling, whereas it remains identically zero for Laplacian diffusion (Fig.~\ref{fig:entropy_rate}\textbf{b}). This association does not imply a general one-to-one correspondence between local instability and positive entropy production, whose sign depends on the instantaneous correlation between the response density and the symmetric part of the Jacobian. Rather, it shows that, in this nonlinear system, $\dot{S}>0$ captures a change in perturbation-response organization occurring across a dynamical stability threshold that the diffusive description cannot reproduce.

\subsection{Informational least-dissipation principle}

Because of the intrinsic limits of the diffusive approach, we develop an information-thermodynamic framework to describe responses to perturbations along general trajectories. We provide a theoretical foundation for the relevant thermodynamic quantities, working for simplicity in the fully dissipative frame; the conservative part can be restored by an appropriate rotation, and the fundamental thermodynamic quantities are invariant under this rotation.

The evolution of all thermodynamic quantities is determined by the flow of the density operator $\boldsymbol{\sigma}$. Thus, we seek a foundational principle from which it can be derived (Appendix~\ref{app::derivation}). For this purpose, we consider the manifold of density matrices endowed with the Bures metric \cite{braunstein1994statistical,dittmann1999explicit}, defining the infinitesimal distance between $\boldsymbol{\sigma}$ and $\boldsymbol{\sigma}+d\boldsymbol{\sigma}$:
\begin{equation}
g_{\boldsymbol{\sigma}}(d\boldsymbol{\sigma}, d\boldsymbol{\sigma})= \operatorname{Tr}(\mathbf{K}_{\boldsymbol{\sigma}}(d\boldsymbol{\sigma})\,d\boldsymbol{\sigma}),
\end{equation}
where $\mathbf{K}_{\boldsymbol{\sigma}}(d\boldsymbol{\sigma})$ satisfies $\boldsymbol{\sigma}\mathbf{K}+\mathbf{K}\boldsymbol{\sigma}=d\boldsymbol{\sigma}$. We interpret this metric as the informational cost of distinguishing neighboring density operators. Unit trace preservation restricts admissible variations to traceless $d\boldsymbol{\sigma}$, equivalently $g_{\boldsymbol{\sigma}}(\boldsymbol{\sigma}, d\boldsymbol{\sigma})=0$. We assume that the system’s intrinsic evolution drives its response to perturbations. Accordingly, we introduce the potential $U(\boldsymbol{\sigma})=-2\operatorname{Tr}(\mathbf{Q}\boldsymbol{\sigma})$, measuring alignment between the intrinsic evolution operator $\mathbf{Q}$ and the density operator. This generalizes the energy-like role that the Laplacian $\mathbf{L}$ has in the diffusive limit. We now state the following variational principle: at any scale $t$ associated with a driving $\mathbf{Q}(t)$, the density operator follows the trace-preserving path with the least informational cost that maximizes the instantaneous alignment with $\mathbf{Q}(t)$. This is obtained by minimizing
\begin{equation}
\mathcal{R}= g_{\boldsymbol{\sigma}}(\dot{\boldsymbol{\sigma}}, \dot{\boldsymbol{\sigma}})+\dot{U}+\alpha \, g_{\boldsymbol{\sigma}}({\boldsymbol{\sigma}}, \dot{\boldsymbol{\sigma}}),
\end{equation}
under $g_{\boldsymbol{\sigma}}({\boldsymbol{\sigma}}, \dot{\boldsymbol{\sigma}})=0$, enforced by the Lagrange multiplier $\alpha$. Imposing $\frac{\delta}{\delta\dot{\boldsymbol{\sigma}}}\mathcal{R}=0$ recovers the density matrix flow equation. The principle determines the dissipative structure, while the conservative part is a gauge choice recoverable by rotating back to the original frame.

\begin{figure*}
\includegraphics[width=\textwidth]{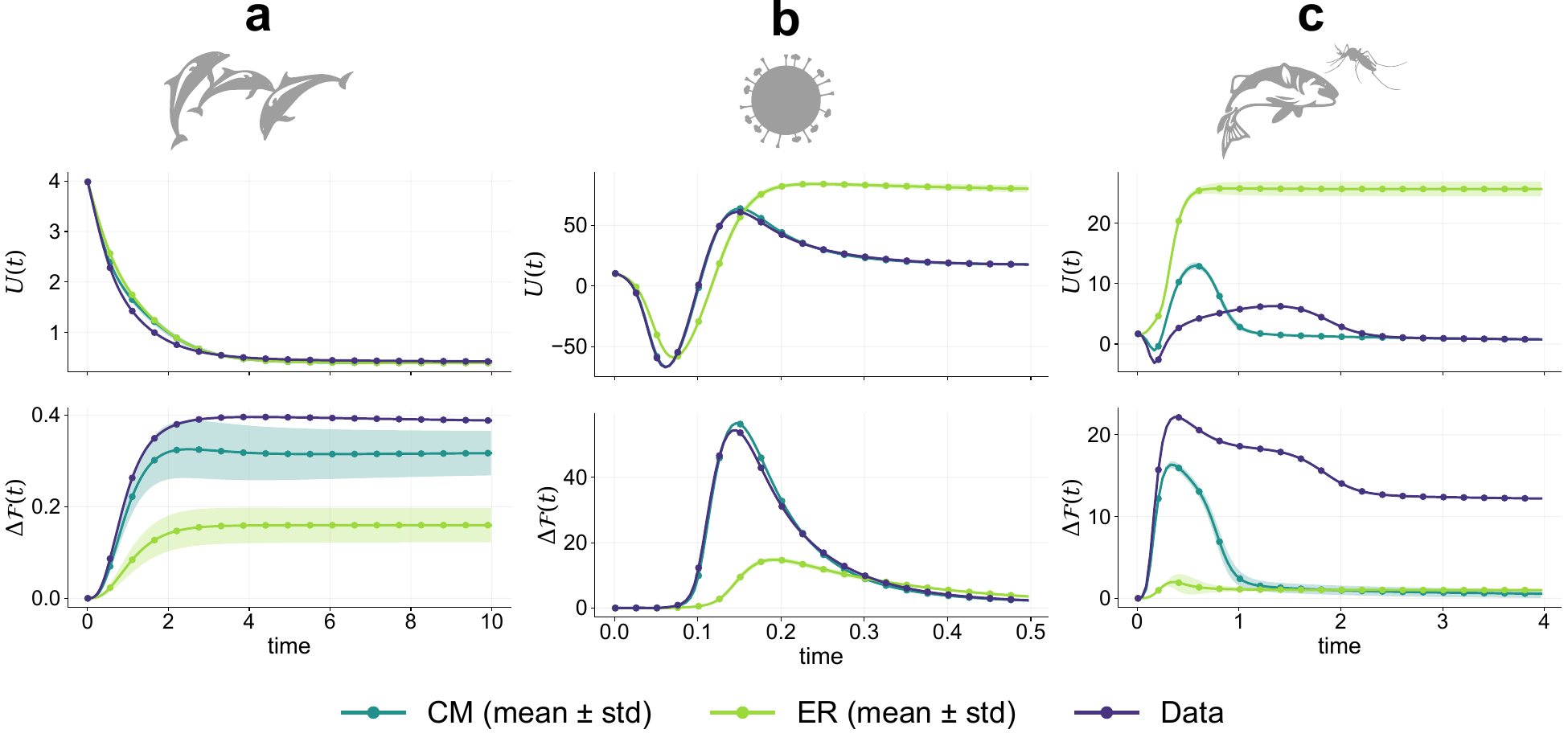}
\caption{\label{fig:free_energy_grid}
\textbf{Time evolution of the informational potential $U(t)$ and of the free energy difference $\Delta \mathcal{F}(t)$ for various dynamical processes on empirical networks and their synthetic model counterparts.} The informational potential $U(t)$ quantifies the average tendency to amplify or curb perturbations, while $\Delta \mathcal{F}(t)=\mathcal{F}-F_{\mathbf{\gamma}}$ measures the distance between a nonlinear system response and its instantaneous diffusive counterpart. We compare these quantities on empirical networks with averages over $100$ configuration model (CM) and Erdős–Rényi (ER) realizations. Noisy voter dynamics ($\mathbb{NV}$) on a dolphin social network \textbf{(a)} has a time-independent, non-symmetric operator $\mathbf{J}$: since $U(t)$ depends only on its symmetric part, the profile is purely contractive and diffusion-like, while the antisymmetric part yields a non-zero distance from diffusion, highlighting the effects of non-normality. SIS dynamics $(\mathbb{E})$ on a contact network \cite{eagle2006reality} \textbf{(b)} and population dynamics $(\mathbb{P})$ on a trophic network \cite{martinez1991artifacts} \textbf{(c)} alternate expansive and contractive phases due to the time dependence of the evolution operator. Overall, empirical networks are often farther from diffusion than synthetic counterparts, as complex structure amplifies the effects of asymmetry, non-normality, and Jacobian time dependence.}
\end{figure*}

This is an informational least dissipation principle, based on the cost of distinguishing neighboring density operators. It is analogous to the Onsager--Rayleigh least dissipation principle \cite{onsager1931reciprocal,onsager1931reciprocal2}, and $\mathcal{R}$ has the same formal structure as the Rayleigh dissipation functional \cite{doi2011onsager}. Moreover, the variational formulation rewrites $\dot{\boldsymbol{\sigma}}$ as a gradient flow driven by the gradient of the time-dependent potential $U$ with respect to the Bures metric (Appendix~\ref{app::gradient}). This supports the extension of $U(t)=-2 \operatorname{Tr}(\mathbf{J}\boldsymbol{\rho})$ as a thermodynamic potential governing the density flow, interpreted as the average contraction ($U>0$) or expansion rate ($U<0$) of perturbations (Fig.~\ref{fig:free_energy_grid}).

\textit{Generalized free energy}\textemdash After the introduction of the generalized framework, we examine its relation to the diffusive approximation at scale $t$. Following non-equilibrium thermodynamics \cite{onsager1931reciprocal,onsager1931reciprocal2}, we combine the potential $U$ with the Von Neumann entropy into the generalized free energy $\mathcal{F}\equiv U-\frac{1}{t}S$. This yields a measure of the distance from the instantaneous diffusive reference $\boldsymbol{\gamma}(t)$, quantifying how well the local perturbation response is approximated by diffusion. Indeed, with some algebra,
\begin{equation}
\mathcal{F}=F_{\boldsymbol{\gamma}}+\frac{1}{t}\mathcal{D}_{KL}(\boldsymbol{\rho}||\boldsymbol{\gamma}),
\end{equation}
where $F_{\boldsymbol{\gamma}}=-\frac{1}{t}\log[\operatorname{Tr}(e^{2t \mathbf{D}})]$ is the Gibbsian free energy of $\boldsymbol{\gamma}$ and $\mathcal{D}_{KL}(\boldsymbol{\rho}||\boldsymbol{\gamma})$ is the Kullback--Leibler divergence between $\boldsymbol{\rho}$ and $\boldsymbol{\gamma}$. Thus, $\mathcal{D}_{KL}$ emerges as a thermodynamically motivated measure of the distance between a system density operator and its Gibbsian approximation.

We evaluate this divergence on real-world networks with appropriate dynamics, comparing them with configuration-model and Erd\H{o}s--R\'enyi counterparts with the same number of nodes and comparable average degree (Fig.~\ref{fig:free_energy_grid}). Two distinct comparisons are involved. For each structure--dynamics pair, $D_{\mathrm{KL}}(\boldsymbol{\rho}\Vert\boldsymbol{\gamma})$ measures the departure of the actual perturbation response from its own instantaneous diffusive reference. These distances can then be compared across network ensembles. Although no strict ordering holds at every time, we observe a recurrent tendency for empirical networks to depart more strongly from their diffusive references than configuration-model networks, and for the latter to depart more strongly than Erd\H{o}s--R\'enyi networks.

This tendency admits a natural information-theoretic interpretation. In maximum-entropy approaches, complexity can be quantified by the information not captured by reconstructions based on a restricted set of constraints~\cite{chliamovitch2014assessing}. The three network classes considered here define an analogous hierarchy of structural information: Erd\H{o}s--R\'enyi networks retain essentially only network size and mean density, the configuration model additionally preserves the degree sequence, while empirical networks retain correlations and higher-order organization removed by randomization. Our framework provides a thermodynamically motivated dynamical counterpart of such complexity measures. Since the response state $\boldsymbol{\rho}(t)$ is generated by $\mathbf{J}[\mathbf{x}(t);\mathbf{A}]$, structural information retained in $\mathbf{A}$ can be reflected in the organization of the perturbation response. Accordingly, $D_{\mathrm{KL}}(\boldsymbol{\rho}\Vert\boldsymbol{\gamma})$ quantifies the response information that is not reducible to the corresponding instantaneous Gibbsian approximation. The recurrent hierarchy in Fig.~\ref{fig:free_energy_grid} therefore suggests that progressively richer structural organization is accompanied, in the systems considered here, by progressively less reducible dynamical responses. We stress that this is an information-theoretic interpretation of the observed statistical tendency rather than as a universal pointwise inequality holding for every realization and time.

\section{Conclusions and outlook}

Real-world complex networks exhibit heterogeneous responses to perturbations. Using information-thermodynamic arguments, we showed that standard diffusive approaches with time-independent evolution operators cannot capture this richness, as they fail to describe transient increases in response diversity. This motivates a trajectory-based formalism covering broad dynamical regimes, including far from steady state ones.

We developed an information-thermodynamic framework based on a least dissipation variational principle, establishing a correspondence with Onsager non-equilibrium thermodynamics analogous to the link between the diffusive-like case and equilibrium thermodynamics. This paves the way for the extension of many tools previously applied successfully in the diffusive-like case to a broader dynamical landscape. Moreover, current renormalization group techniques for complex networks \cite{villegas2023laplacian,nurisso2025higher} are based on the preservation of the dissipative structure of Laplacian diffusion across time scales. Our formalism manages to encode the dissipative structure of many dynamics at any scale $t$ in $\mathcal{R}(t)$, opening the possibility for a network renormalization group that incorporates both structural and dynamical features governing information spreading: a key step toward a general renormalization theory for interacting complex systems.

The normalization-preserving factor in the flow equation is consistent with analogous terms in probability-density equations defined on systems whose intrinsic dynamics are not probability preserving, such as the classical Liouville equation for non-Hamiltonian systems \cite{j2007statistical} or the Lindblad equation for open quantum systems \cite{manzano2020short}. A structure combining intrinsic evolution with a normalization term linear in the distribution and proportional to the generator average is also typical of replicator dynamics \cite{page2002unifying}. Thus, equation (\ref{Liouville}) can be viewed as a generalized matrix replicator equation \cite{saakian2004eigen}, where $\mathbf{J}(\mathbf{x}(t))$ acts as a fitness matrix for $\boldsymbol{\rho}$, aligning perturbation ensembles with instantaneous evolution directions. These links may connect the framework to Bayesian inference \cite{jaynes2003probability} and support new network inference techniques.

\paragraph{Author contributions.} A.F.B. and M.D.D. designed the study, performed the theoretical analysis and wrote the manuscript. A.F.B. performed the numerical experiments.

\paragraph{Competing interests.} The authors declare no competing interests.


\begin{appendices}

\numberwithin{equation}{section}
\section{Response to perturbations of a dynamical system} \label{app::response}
We consider a complex system composed of a set of $N$ interacting units, or nodes. At any time $t$, each node $i$ is associated with a state variable $x_i(t)$, and the collective state of the system is represented by the $N$-dimensional vector $\ket{\mathbf{x}(t)}$. The time evolution of the system is determined by both its structure and its dynamics. Structural interactions are encoded in an adjacency matrix $\mathbf{A}$, whose entry $A_{ij}$ represents the strength of the interaction from node $j$ to node $i$, while the dynamics are specified by a generally nonlinear vector field $\mathbf{v}(\ket{\mathbf{x}(t)};\mathbf{A})$.

Throughout this letter, we focus on systems governed by first-order differential equations,
$\frac{d}{dt}\ket{\mathbf{x}(t)}=\mathbf{v}(\ket{\mathbf{x}(t)};\mathbf{A}) .$
To probe the system’s sensitivity to perturbations, we consider a small displacement $\ket{\delta\mathbf{x}(t)}$ of the state vector. Linearizing the dynamics around the trajectory $\ket{\mathbf{x}(t)}$, the evolution of the perturbation is governed by the Jacobian matrix $\mathbf{J}$, with elements $J_{ij}=\frac{\partial v_i}{\partial x_j}$, according to
\begin{equation}\label{pert_ev}
\frac{d}{dt}\ket{\delta \mathbf{x}(t)}=\mathbf{J}(\mathbf{x}(t))\ket{\delta \mathbf{x}(t)}.
\end{equation}
The Jacobian is evaluated along the trajectory generated by $\mathbf{v}(\ket{\mathbf{x}(t)};\mathbf{A})$ together with the chosen initial condition. Our aim is to characterize the response to perturbations of a complex system along any trajectory generated by $\mathbf{v}(\ket{\mathbf{x}(t)};\mathbf{A})$, given an initial condition $\ket{\mathbf{x}(t_0)}=\ket{\mathbf{x}_0}$.

\section{Correspondence between the diffusive case and equilibrium statistical mechanics} \label{app::correspondence}
For diffusive dynamics, the evolution of the system is determined by a linear vector field,
$\mathbf{v}(\ket{\mathbf{x}(t)};\mathbf{A})= -\mathbf{L}\ket{\mathbf{x}(t)},$
where, in the common case of Laplacian diffusion, $\mathbf{L}$ is a time-independent Laplacian matrix~\cite{masuda2017random}, with entries $L_{ij}=k_i \delta_{ij}-A_{ij}$ and $k_i$ denoting the degree of node $i$. In the case of Laplacian diffusion on an undirected network, equation~(\ref{pert_ev}) can then be solved with constant $\mathbf{J}=-\mathbf{L}$, yielding the density matrix at any propagation scale $t$ as
\begin{equation}\label{diff_dm}
    \boldsymbol{\rho}(t)=\frac{e^{-2t \mathbf{L}}}{\operatorname{Tr}(e^{-2t \mathbf{L}})}.
\end{equation}

Formally, the density matrix defined in equation~(\ref{diff_dm}) is equivalent to a canonical Gibbs state associated with the network Laplacian~\cite{landau2013statistical}, with $\mathbf{L}$ playing the role of a Hamiltonian-like operator and $t$ acting as a scale parameter, as the classical inverse temperature $\beta$ of equilibrium statistical mechanics. This correspondence establishes a direct link between diffusion on networks and equilibrium statistical mechanics, enabling the construction of an information--thermodynamic framework for diffusive dynamics~\cite{ghavasieh2022statistical}. In this framework, $\boldsymbol{\rho}$ acts as a density operator and $Z$ plays the role of a partition function, from which a free energy
$F=-\frac{1}{t}\log(Z)$
can be defined. This quantity has been interpreted as a measure of the speed of information propagation across the network~\cite{ghavasieh2020enhancing}.

The associated Von Neumann entropy,
$S=-\operatorname{Tr}(\boldsymbol{\rho}\log\operatorname{\boldsymbol{\rho}}),$
quantifies how evenly perturbation modes contribute to the system’s response. High entropy corresponds to a broad distribution of contributing modes, while low entropy indicates the dominance of a small number of modes. The entropy is maximized when all modes contribute equally, yielding $S=\log N$, which provides a natural normalization such that the normalized entropy satisfies $S\in[0,1]$. Within this interpretation, the entropy captures the functional diversity of the system’s response to perturbations.

The free energy $F$, entropy $S$, and Laplacian $\mathbf{L}$ satisfy the information-theoretic relation~\cite{amari2016information}
\begin{equation}\label{F_def}
F=U-\frac{1}{t}S,
\end{equation}
where
$U=2 \operatorname{Tr}(\mathbf{L}\boldsymbol{\rho})$
defines an average internal energy consistent with the role of $\mathbf{-L}$ as an evolution operator. At the scale $t=0$, this energy has a clear physical interpretation as it is proportional to the number of edges in the network.

\section{Density flow equation} \label{app::density}
Given an ensemble of orthogonal perturbations $\Psi_{t}= \{\delta \mathbf{\mathbf{x}}^\alpha (t)\}_{\alpha=0,...,N-1}$, we build the associated density matrix as 
\begin{equation} \label{rho_def}
    \boldsymbol{\rho}(t)=\sum_\alpha p_\alpha \frac{\ket{\delta \mathbf{x}^\alpha(t)}\bra{\delta \mathbf{x}^\alpha(t)}}{Z(t)}=\frac{\boldsymbol{\mathcal{U}}(t)}{Z(t)},
\end{equation}
where the $\{p_\alpha\}_{\alpha=0,...,N-1}$ are the initial probabilities assigned to each perturbation mode and 
\begin{equation}
   Z(t)=\operatorname{Tr}\left(\sum_\alpha p_\alpha\ket{\delta \mathbf{x}^\alpha(t)}\bra{\delta \mathbf{x}^\alpha(t)}\right). 
\end{equation}
By using (\ref{pert_ev}), we derive that
\begin{equation} \label{prop_ev}
    \frac{d}{dt}\boldsymbol{\mathcal{U}}=\mathbf{J}\boldsymbol{\mathcal{U}}+\boldsymbol{\mathcal{U}}\mathbf{J}^T
\end{equation}
and
\begin{equation} \label{invZ_ev}
    \frac{d}{dt}\frac{1}{Z}=-\frac{1}{Z^2}[2\operatorname{Tr}(\boldsymbol{\mathcal{U}}\mathbf{J})]=-2\frac{\operatorname{Tr}(\boldsymbol{\rho}\mathbf{J})}{Z}.
\end{equation}
Then, exploiting (\ref{prop_ev}) and (\ref{invZ_ev}), we directly recover the flow equation by applying the time derivative to the definition in (\ref{rho_def}):
\begin{equation}
    \dot{\boldsymbol{\rho}}=\mathbf{J} \boldsymbol{\rho}+ \boldsymbol{\rho} \mathbf{J}^T-2 \boldsymbol{\rho} \operatorname{Tr}(\boldsymbol{\rho}\mathbf{J}).
\end{equation}

Now, consider an ensemble of perturbations $\Psi_{t_0}= \{\delta \mathbf{\mathbf{x}}^\alpha (t_0)\}_{\alpha=0,...,N-1}$ that is an orthonormal basis at time $t_0$. We associate to it a density operator $\boldsymbol{\rho}(t_0)$, whose evolution is given by equation (\ref{Liouville}). Given the initial perturbation density operator $\boldsymbol{\rho}(t_0)$, the solution of equation (\ref{Liouville}) is 
\begin{equation} \label{solution}
    \boldsymbol{\rho}(t)=\frac{\mathbf{\Phi}(t,t_0)\boldsymbol{\rho}(t_0)\mathbf{\Phi}(t,t_0)^T}{\operatorname{Tr}[\mathbf{\Phi}(t,t_0)\boldsymbol{\rho}(t_0)\mathbf{\Phi}(t,t_0)^T]},
\end{equation}
in which $\mathbf{\Phi}(t,t_0)= \mathcal{T} \{ \exp[\int_{t_0}^{t}ds\, \mathbf{J}(s)]\}$ is the time-ordered exponential built from the Jacobian, and $\mathbf{\Phi}^T$ is the anti-time-ordered exponential built from its transpose. This general solution yields a density matrix that is a matrix functional $\boldsymbol{\rho}[\mathbf{x}(t);t]$ of the trajectory $\mathbf{x}(t)$ and allows for the direct extension of many of the applications previously cited for the diffusive case to the general one.

We want to investigate how the symmetric part $\boldsymbol{D}$ and the antisymmetric part $\boldsymbol{C}$ of $\boldsymbol{J}$ influence the evolution of the eigenvalues and eigenvectors of (\ref{solution}). We begin by substituting this separation into the flow equation, obtaining
\begin{equation}
    \dot{\boldsymbol{\rho}}=\{\mathbf{D},\boldsymbol{\rho}\}-2 \boldsymbol{\rho} \operatorname{Tr}(\boldsymbol{\rho}\mathbf{D})+[\mathbf{C},\boldsymbol{\rho}].
\end{equation}
 According to Rayleigh's matrix perturbation formulas, given a matrix $\boldsymbol{M}$ with starting eigenvalues $m_i$ and associated eigenvectors $\ket{m_i}$, the shifted matrix $\boldsymbol{B}=\boldsymbol{M}+\delta \boldsymbol{M}$ has eigenvalues 
 \begin{equation}
     b_i=m_i+\bra{m_i} \delta \boldsymbol{M} \ket{m_i},
 \end{equation}
and eigenvectors
\begin{equation}
    \ket{b_i}=\ket{m_i}+\sum_{j\neq i}\frac{\bra{m_j} \delta \boldsymbol{M} \ket{m_i}}{m_i - m_j}\ket{m_j},
\end{equation}
for a small enough shift $\delta \boldsymbol{M}$. We can apply these formulas to the density matrix evolution considering the shift $\delta \boldsymbol{\rho}= \delta t \frac{d \boldsymbol{\rho}}{dt}$, and substitute the explicit form of the flow equation. By doing so and indicating with $p_i(t)$ and $\ket{p_i(t)}$ respectively the eigenvalues and the eigenvectors of $\boldsymbol{\rho}$ at time $t$, we get for the eigenvalues
\begin{equation}
    p_i(t+\delta t)=p_i(t)\big[1+2 \delta t \big( \bra{p_i(t)} \boldsymbol{D} \ket{p_i(t)}-\operatorname{Tr}(\boldsymbol{\rho}\boldsymbol{D}) \big) \big]
\end{equation}
and, for the eigenvectors,
\begin{equation}
    \ket{p_i(t+\delta t)}=\ket{p_i(t)}+\delta t \sum_{j\neq i} \big[\frac{p_i(t) + p_j(t)}{p_i(t) - p_j(t)}\bra{p_j(t)} \boldsymbol{D} \ket{p_i(t)}+\bra{p_j(t)} \boldsymbol{C} \ket{p_i(t)}\big]\ket{p_j(t)}.
\end{equation}
This shows that the symmetric part $\boldsymbol{D}$ affects both the eigenvalues and eigenvectors, while the antisymmetric part $\boldsymbol{C}$ only acts on the eigenvectors, generating a unitary rotational evolution.

\section{Network H-theorem} \label{app::network}
In this section, we characterize the conditions under which the response entropy is guaranteed to be non-increasing. For a density operator $\boldsymbol{\rho}(t)$ evolving according to the previously discussed flow equation, the entropy rate is
\begin{equation}\label{eq:H_cov}
    \dot{S}= -2 \,\operatorname{Cov}_{\boldsymbol{\rho}}(\mathbf{D},\log\boldsymbol{\rho}).
\end{equation}
Therefore, at any given time, the general instantaneous condition to have a non-increasing entropy is
\begin{equation}\label{eq:H_local_condition}
    \dot S(t)\leq 0 \quad\Longleftrightarrow \quad \operatorname{Cov}_{\boldsymbol{\rho}}(\mathbf{D},\log\boldsymbol{\rho}) \geq 0.
\end{equation}
In particular, the sign of $\dot S$ is not determined by the instantaneous symmetric operator $\mathbf{D}(t)$ alone, but by its relation to the current density operator $\boldsymbol{\rho}(t)$, which contains the accumulated effect of both the initial perturbation ensemble and the preceding dynamics.

The physical content of the $H$-theorem lies in identifying classes of evolution for which the features of the underlying dynamics and of the chosen prior always forbid an increase of entropy. A relevant class of this kind is represented by dynamics with a time-independent normal evolution operator $\mathbf{J}$ starting from a maximum entropy prior $\boldsymbol{\rho}(0)=\mathbf{I}/N$. Indeed, in this case,
\begin{equation}\label{eq:normal_rho}
    \boldsymbol{\rho}(t)
    =\frac{e^{\mathbf{J}t}e^{\mathbf{J}^{T}t}}{\mathrm{Tr}\!\left(e^{\mathbf{J}t}e^{\mathbf{J}^{T}t}\right)}=\frac{e^{2t\mathbf{D}}}{\mathrm{Tr}\!\left(e^{2t\mathbf{D}}\right)},
\end{equation}
hence,
\begin{equation}
    \log\boldsymbol{\rho}(t)=2t\mathbf{D}-\log Z(t)\,\mathbf{I},
\end{equation}
and Eq.~\eqref{eq:H_cov} gives
\begin{equation}\label{eq:H_normal}
    \dot S(t)=-4t\,\mathrm{Var}_{\boldsymbol{\rho}(t)}(\mathbf{D})\leq 0.
\end{equation}
The time-independent symmetric diffusive case discussed in the main text corresponds to the special case $\mathbf{J}=\mathbf{D}$. By imposing a few additional constraints on this class, the theorem can be extended to dynamics governed by a symmetric time-dependent evolution operator. In particular, we require a symmetric time-dependent operator $\mathbf{D}(t)$ satisfying
\begin{equation}
    [\mathbf{D}(t),\mathbf{D}(s)]=0
    \qquad
    \forall\,s,t,
\end{equation}
and assume that the ordering of its eigenvalues,
\begin{equation}
    d_1(t)\geq d_2(t)\geq\cdots\geq d_N(t),
\end{equation}
is preserved throughout the interval of interest. In this case, starting from $\boldsymbol{\rho}(0)=\boldsymbol{I}/N$ and defining the accumulated stretching rates as
\begin{equation}
    A_i(t)=\int_0^t d_i(s)\,ds,
\end{equation}
the eigenvalues of the density operator are
\begin{equation}
    p_i(t)=\frac{\exp\!\left[2A_i(t)\right]}{\sum_j \exp\!\left[2A_j(t)\right]} .
\end{equation}
Since the ordering of the instantaneous stretching rates is preserved, it follows that
\begin{equation}
    d_i(t)\ge d_j(t)\quad\Longrightarrow\quad A_i(t)\ge A_j(t)\qquad (i<j).
\end{equation}
Writing the covariance in equation (\ref{eq:H_cov}) in the common eigenbasis of $\mathbf{D}(t)$ and $\boldsymbol{\rho}(t)$, and using $\log p_i(t)=2A_i(t)-\log Z(t)$, we obtain
\begin{equation}
    \dot S(t)=-4\left[\sum_i p_i(t)\,d_i(t)\,A_i(t)-\left(\sum_i p_i(t)\,d_i(t)\right)\left(\sum_j p_j(t)\,A_j(t)\right)\right].
\end{equation}
Using $\sum_i p_i=1$ and symmetrizing, this becomes
\begin{equation}
    \dot S(t)=-4\sum_{i<j}p_i(t)p_j(t)\left[d_i(t)-d_j(t)\right]\left[A_i(t)-A_j(t)\right]\le 0.
\end{equation}
Indeed, each term in the sum is non-negative because the instantaneous and accumulated stretching rates have the same ordering. Therefore, the entropy remains non-increasing throughout any interval over which this ordering is preserved.

Aside from the theorem's range of applicability, it is important to investigate its limitations and the mechanisms that may induce $\dot{S}(t)>0$. In the following sections, we therefore discuss explicitly how entropy can increase when the assumptions on the initial perturbation ensemble or the evolution operator are relaxed. Specifically, we start from the time-independent, symmetric diffusive case with a maximum-entropy prior and examine separately the effects of introducing a biased prior, non-normality, and time dependence, showing through examples how each of these elements can induce time intervals with $\dot{S}>0$. Indeed, the examples will identify three distinct mechanisms that influence the sign of the entropy derivative. A non-uniform prior changes the initial relation between the density operator and the stretching directions even for ordinary diffusion. Time dependence can reverse the accumulated concentration of response modes even when the instantaneous evolution operator is symmetric. Non-normality can dynamically rotate response directions relative to the corresponding symmetric stretching operator even for a constant evolution generator.

\section{Dependence on the perturbation prior} \label{app::prior}
The maximum-entropy prior is essential to the monotonicity result that we obtain in the main text for the standard diffusive case. To show this, we consider the time-independent symmetric diffusion operator $\mathbf{J}=\mathbf{L}$, but choose the non-uniform initial density
\begin{equation}\label{eq:biased_prior}
    \boldsymbol{\rho}(0) =\frac{e^{c \mathbf{L}}}{\mathrm{Tr}\!\left(e^{c\mathbf{L}}\right)},\qquad c>0.
\end{equation}
Since $\boldsymbol{\rho}(0)$ commutes with $\mathbf{L}$, its evolution is
\begin{equation}\label{eq:biased_prior_evolution}
    \boldsymbol{\rho}(t) = \frac{e^{-t\mathbf{L}}\rho(0)e^{-t\mathbf{L}}}{\mathrm{Tr}\!\left[e^{-t\mathbf{L}}\rho(0)e^{-t\mathbf{L}}\right]}=\frac{e^{(c-2t)\mathbf{L}}}{\mathrm{Tr}\!\left[e^{(c-2t)\mathbf{L}}\right]}.
\end{equation}
Therefore,
\begin{equation}
    \log\boldsymbol{\rho}(t) = (c-2t)\mathbf{L}-\log Z(t)\,\mathbf{I}.
\end{equation}
Since $\mathbf{D}=-\mathbf{L}$, Eq.~\eqref{eq:H_cov} becomes
\begin{equation}\label{eq:biased_prior_entropy}
    \dot S(t) =  2(c-2t)\,\mathrm{Var}_{\boldsymbol{\rho}(t)}(\mathbf{L}).
\end{equation}
For every nontrivial $\mathbf{L}$, it follows that
\begin{equation}
    \dot S(t)>0,
    \qquad
    0<t<\frac{c}{2}.
\end{equation}
At $t=c/2$, $\boldsymbol{\rho}=\boldsymbol{I}/N$ and the entropy reaches its maximum, after which the usual diffusive entropy decrease is recovered.

Thus, diffusion alone does not determine the sign of the entropy rate. A biased perturbation prior may initially evolve toward a more uniform response distribution and therefore produce $\dot S>0$ even under a constant symmetric diffusion operator. The maximum-entropy prior adopted in the main text removes this independent source of directional bias and allows us to isolate the effect of the evolution operator.

\section{Time dependence} \label{app::time}
Time dependence provides a distinct mechanism for breaking entropy monotonicity, even when the evolution operator is symmetric and commutes with itself at different times. To see this, consider
\begin{equation}\label{eq:time_dep_D}
    \mathbf{J}(t)=\mathbf{D}(t)
    =
    a(t)
    \begin{pmatrix}
        1&0\\
        0&-1
    \end{pmatrix},
    \qquad
    a(t)=1-2t,
    \qquad
    0\leq t\leq1.
\end{equation}
The generator is symmetric for every $t$, and
\begin{equation}
    [\mathbf{D}(t),\mathbf{D}(s)]=0
    \qquad
    \forall\,s,t.
\end{equation}
Starting again from $\boldsymbol{\rho}(0)=\mathbf{I}/2$, define
\begin{equation}
    A(t)
    =
    \int_0^t a(s)\,ds
    =
    t-t^2.
\end{equation}
The density operator becomes
\begin{equation}\label{eq:time_dep_rho}
    \boldsymbol{\rho}(t)
    =
    \frac{
    \exp\!\left[2A(t)\boldsymbol{\sigma}_z\right]
    }{
    \mathrm{Tr}\exp\!\left[2A(t)\boldsymbol{\sigma}_z\right]
    },
\end{equation}
where $\boldsymbol{\sigma}_z=\mathrm{diag}(1,-1)$. Its entropy is
\begin{equation}
    S(t)
    =
    \log\!\left[2\cosh\!\left(2A(t)\right)\right]
    -
    2A(t)\tanh\!\left(2A(t)\right),
\end{equation}
and therefore
\begin{equation}
    \dot S(t)
    =
    -4A(t)a(t)\,
    \mathrm{sech}^{2}\!\left(2A(t)\right).
    \label{eq:time_dep_entropy}
\end{equation}
Since $A(t)>0$ for $0<t<1$,
\begin{equation}
    \dot S(t)<0,
    \qquad
    0<t<\frac{1}{2},
\end{equation}
whereas
\begin{equation}
    \dot S(t)>0,
    \qquad
    \frac{1}{2}<t<1.
    \label{eq:time_dep_positive}
\end{equation}
At $t=1$, $A(1)=0$ and the maximum-entropy state is recovered.

This example also clarifies the conditions required for a time-dependent extension of the $H$-theorem. Pairwise commutation and a common eigenbasis are not sufficient if the ordering of the instantaneous stretching rates changes during the evolution.

\section{Non-normality of the evolution operator} \label{app::normality}
Time dependence of the evolution operator and a biased prior are not required to obtain a positive entropy derivative; indeed, non-normality can also generate intervals with $\dot{S}>0$. To isolate its effect, consider the constant two-dimensional linear evolution operator
\begin{equation}\label{eq:nonnormal_J}
    \mathbf{J}=-\gamma \mathbf{I}+\mathbf{A},
    \qquad
    \mathbf{A}=
    \begin{pmatrix}
        0 & -\alpha\\
        \alpha^{-1} & 0
    \end{pmatrix},
    \qquad
    \alpha>1.
\end{equation}
For $\alpha\neq1$,
\begin{equation}
    \mathbf{A}\mathbf{A}^{T}\neq \mathbf{A}^{T}\mathbf{A},
\end{equation}
so that $\mathbf{J}$ is non-normal. The optional isotropic contribution $-\gamma \mathbf{I}$, with $\gamma>0$, makes the linear dynamics asymptotically stable, but cancels out of the normalized density operator and therefore does not affect $S(t)$.

Since
\begin{equation}
    \mathbf{A}^2=-\mathbf{I},
\end{equation}
we have
\begin{equation}
    e^{\mathbf{A}t}
    =
    (\cos t)\,\mathbf{I}+(\sin t)\,\mathbf{A}.
\end{equation}
Starting from the maximum-entropy state $\boldsymbol{\rho}(0)=\mathbf{I}/2$, let
\begin{equation}
    \boldsymbol{\Phi}(t)=e^{\mathbf{A}t}.
\end{equation}
The density operator is then
\begin{equation}\label{eq:nonnormal_rho}
    \boldsymbol{\rho}(t)
    =
    \frac{\boldsymbol{\Phi}(t)\boldsymbol{\Phi}^{T}(t)}
    {\mathrm{Tr}\!\left[\boldsymbol{\Phi}(t)\boldsymbol{\Phi}^{T}(t)\right]}.
\end{equation}
The determinant of $\boldsymbol{\Phi}(t)\boldsymbol{\Phi}^{T}(t)$ is unity, while its trace is
\begin{equation}\label{eq:nonnormal_trace}
    \tau(t)\equiv \operatorname{Tr}[ \boldsymbol{\Phi}(t)\boldsymbol{\Phi}^{T}(t)]
    =
    2+
    \left(\alpha-\alpha^{-1}\right)^2
    \sin^2 t,
\end{equation}
from which it follows that $\tau(t)\geq2$.
The two eigenvalues of $\boldsymbol{\rho}(t)$ are therefore
\begin{equation}\label{eq:nonnormal_eigenvalues}
    p_{\pm}(t)
    =
    \frac{1}{2}
    \left[
    1\pm
    \sqrt{1-\frac{4}{\tau^2(t)}}
    \right],
\end{equation}
and the entropy is
\begin{equation}
    S(t)
    =
    -p_{+}\log p_{+}
    -p_{-}\log p_{-}.
\end{equation}
Using the chain rule, we write the time derivative of the entropy as
\begin{equation}
    \dot{S}= \big(\frac{\partial S}{\partial p_+}\big) \,  \big (\frac{\partial p_+}{\partial \tau} \big)\; \dot{\tau},
\end{equation}
and observe that its sign can change only through the factor $\dot{\tau}$. Indeed, whenever $\dot{S}\neq0$, and hence $\tau>2$, we have $\frac{\partial S}{\partial p_+}<0$ and $\frac{\partial p_+}{\partial \tau}>0$.
Thus, since
\begin{equation}
    \dot\tau(t)
    =
    \left(\alpha-\alpha^{-1}\right)^2
    \sin(2t),
\end{equation}
the entropy decreases for $0<t<\pi/2$, but increases for
\begin{equation}\label{eq:nonnormal_positive_entropy}
    \frac{\pi}{2}<t<\pi,
    \qquad
    \dot S(t)>0.
\end{equation}
At $t=\pi$, $\boldsymbol{\Phi}=-\mathbf{I}$, and the maximum-entropy state $\boldsymbol{\rho}=\mathbf{I}/2$ is recovered.

This example shows that a system with a constant evolution operator and starting from the same maximum-entropy prior used in the main text can exhibit finite intervals of increasing response entropy solely as a consequence of non-normality. In terms of the decomposition $\mathbf{J}=\mathbf{D}+\mathbf{C}$, the antisymmetric contribution $\mathbf{C}$ does not directly change the eigenvalues of $\boldsymbol{\rho}$ at a given instant. For a non-normal generator, however, $[\mathbf{D},\mathbf{C}]\neq0$, and the rotation induced by $\mathbf{C}$ continuously changes the orientation of the response eigenvectors relative to the stretching directions of $\mathbf{D}$. This modifies their subsequent evolution and can reverse the sign of the covariance in Eq.~\eqref{eq:H_cov}.

\section{Derivation of the flow equation from the least dissipation principle} \label{app::derivation}
We consider the manifold $\mathcal{D}$ of full-rank density operators $\boldsymbol{\sigma}$ with
$\operatorname{Tr}\boldsymbol{\sigma}=1$, endowed with the Bures metric. Given a tangent vector $\boldsymbol{X}$ at $\boldsymbol{\sigma}$, we define the operator $\boldsymbol{K}_{\boldsymbol{\sigma}}(\boldsymbol{X})$ as the unique solution of the equation
\begin{equation}
\boldsymbol{\sigma}\,\boldsymbol{K}_{\boldsymbol{\sigma}}(\boldsymbol{X})
+
\boldsymbol{K}_{\boldsymbol{\sigma}}(\boldsymbol{X})\,\boldsymbol{\sigma}
=
\boldsymbol{X}.
\label{eq:sylvester}
\end{equation}
Given two tangent vectors $\mathbf{X}$ and $\mathbf{Y},$ the Bures metric is then defined as:
\begin{equation}
g_{\boldsymbol{\sigma}}(\boldsymbol{X},\boldsymbol{Y})
=
\operatorname{Tr}\!\big(
\boldsymbol{K}_{\boldsymbol{\sigma}}(\boldsymbol{X})\,\boldsymbol{Y}
\big),
\label{eq:bures_metric}
\end{equation}
which is symmetric and positive definite on the space of trace-preserving variations. Since $\operatorname{Tr}\boldsymbol{\sigma}=1$ at any scale $t$, admissible velocities satisfy
$\operatorname{Tr}\dot{\boldsymbol{\sigma}}=0$. Using \eqref{eq:sylvester} with $\boldsymbol{X}=\boldsymbol{\sigma}$, we find $\boldsymbol{K}_{\boldsymbol{\sigma}}(\boldsymbol{\sigma})=\tfrac12 \boldsymbol{I}$, hence we write this constraint as
\begin{equation}
g_{\boldsymbol{\sigma}}(\boldsymbol{\sigma},\dot{\boldsymbol{\sigma}})
=
\frac12\,\operatorname{Tr}(\dot{\boldsymbol{\sigma}})
=
0.
\label{eq:trace_constraint}
\end{equation}
Moreover, varying the tangent vector at fixed $\boldsymbol{\sigma}$, we have
\begin{equation}
    \delta g_{\boldsymbol{\sigma}}(\boldsymbol{X},\boldsymbol{X})
= \operatorname{Tr}\!\big(
\delta \boldsymbol{K}_{\boldsymbol{\sigma}}(\boldsymbol{X})\,\boldsymbol{X}
\big)+\operatorname{Tr}\!\big(
\boldsymbol{K}_{\boldsymbol{\sigma}}(\boldsymbol{X})\,\delta\boldsymbol{X}
\big).
\end{equation}
Then, exploiting \eqref{eq:sylvester}, we write
\begin{equation}
    \operatorname{Tr}\!\big(
\delta \boldsymbol{K}_{\boldsymbol{\sigma}}(\boldsymbol{X})\,\boldsymbol{X}
\big)=\operatorname{Tr}\!\big(
\delta \boldsymbol{K}_{\boldsymbol{\sigma}}(\boldsymbol{X})\,\boldsymbol{\sigma}\,\boldsymbol{K}_{\boldsymbol{\sigma}}(\boldsymbol{X})\big)+\operatorname{Tr}\!\big(
\delta \boldsymbol{K}_{\boldsymbol{\sigma}}(\boldsymbol{X})\,\boldsymbol{K}_{\boldsymbol{\sigma}}(\boldsymbol{X})\,\boldsymbol{\sigma}\big).
\label{eq:passage}
\end{equation}
Applying the cyclic property of the trace to the second term of the right hand side of this equation, we get to
\begin{equation}
    \operatorname{Tr}\!\big(
\delta \boldsymbol{K}_{\boldsymbol{\sigma}}(\boldsymbol{X})\,\boldsymbol{K}_{\boldsymbol{\sigma}}(\boldsymbol{X})\,\boldsymbol{\sigma}\big)=\operatorname{Tr}\!\big(\boldsymbol{\sigma}
\,\delta \boldsymbol{K}_{\boldsymbol{\sigma}}(\boldsymbol{X})\,\boldsymbol{K}_{\boldsymbol{\sigma}}(\boldsymbol{X})\big)
\label{eq:cyclic}
\end{equation}
Subsequently, noticing that $\boldsymbol{\sigma}\,\delta\boldsymbol{K}_{\boldsymbol{\sigma}}(\boldsymbol{X})
+
\delta \boldsymbol{K}_{\boldsymbol{\sigma}}(\boldsymbol{X})\,\boldsymbol{\sigma}
=
\delta \boldsymbol{X}$ and considering together \eqref{eq:passage} and \eqref{eq:cyclic}, we obtain
\begin{equation}
   \operatorname{Tr}\!\big(
\delta \boldsymbol{K}_{\boldsymbol{\sigma}}(\boldsymbol{X})\,\boldsymbol{X}
\big) =\operatorname{Tr}\!\big(
\boldsymbol{K}_{\boldsymbol{\sigma}}(\boldsymbol{X})\,\delta\boldsymbol{X}
\big),
\end{equation}
which yields $\delta g_{\boldsymbol{\sigma}}(\boldsymbol{X},\boldsymbol{X})
=2\operatorname{Tr}\!\big(
\boldsymbol{K}_{\boldsymbol{\sigma}}(\boldsymbol{X})\,\delta\boldsymbol{X}
\big)$ and thus
\begin{equation}
    \frac{\delta g_{\boldsymbol{\sigma}}(\boldsymbol{X},\boldsymbol{X})}{\delta\boldsymbol{X}}=2\boldsymbol{K}_{\boldsymbol{\sigma}}(\boldsymbol{X}).
\end{equation}
We introduce the informational potential
\begin{equation}
U(\boldsymbol{\sigma},t)
=
-2\,\operatorname{Tr}\!\big(\boldsymbol{Q}(t)\boldsymbol{\sigma}\big),
\label{eq:potential}
\end{equation}
and the dissipation functional
\begin{equation}
\mathcal{R}
=
g_{\boldsymbol{\sigma}}(\dot{\boldsymbol{\sigma}},\dot{\boldsymbol{\sigma}})
+
\dot U
+
\alpha\,g_{\boldsymbol{\sigma}}(\boldsymbol{\sigma},\dot{\boldsymbol{\sigma}}),
\label{eq:rayleighian}
\end{equation}
where the term with $\alpha$ enforces the constraint \eqref{eq:trace_constraint}
and $\dot U = -2\,\operatorname{Tr}\!\big(\boldsymbol{Q}\,\dot{\boldsymbol{\sigma}}\big)-2\,\operatorname{Tr}\!\big(\dot{\boldsymbol{Q}}\,\boldsymbol{\sigma}\big)$. Varying \eqref{eq:rayleighian} with respect to $\dot{\boldsymbol{\sigma}}$
at fixed $(\boldsymbol{\sigma},t)$ and then imposing $\frac{\delta}{\delta\dot{\boldsymbol{\sigma}}}\mathcal{R}=0$, yields
\begin{equation}
2\,\boldsymbol{K}_{\boldsymbol{\sigma}}(\dot{\boldsymbol{\sigma}})
-
2\,\boldsymbol{Q}
+
\frac{\alpha}{2}\boldsymbol{I}
=
0,
\end{equation}
or equivalently
\begin{equation}
\boldsymbol{K}_{\boldsymbol{\sigma}}(\dot{\boldsymbol{\sigma}})
=
\boldsymbol{Q}
-
\frac{\alpha}{4}\boldsymbol{I}.
\label{eq:Ksigmadot}
\end{equation}
Using \eqref{eq:sylvester} with $\boldsymbol{X}=\dot{\boldsymbol{\sigma}}$
and \eqref{eq:Ksigmadot},
\begin{equation}
\dot{\boldsymbol{\sigma}}
=
\boldsymbol{\sigma}\boldsymbol{K}_{\boldsymbol{\sigma}}(\dot{\boldsymbol{\sigma}})
+
\boldsymbol{K}_{\boldsymbol{\sigma}}(\dot{\boldsymbol{\sigma}})\boldsymbol{\sigma}
=
\{\boldsymbol{Q},\boldsymbol{\sigma}\}
-
\frac{\alpha}{2}\boldsymbol{\sigma}.
\label{eq:flow_prealpha}
\end{equation}
Imposing $\operatorname{Tr}\dot{\boldsymbol{\sigma}}=0$ fixes
$\alpha=4\,\operatorname{Tr}(\boldsymbol{Q}\boldsymbol{\sigma})$, leading to
\begin{equation}
\boxed{
\dot{\boldsymbol{\sigma}}
=
\{\boldsymbol{Q},\boldsymbol{\sigma}\}
-
2\,\operatorname{Tr}(\boldsymbol{Q}\boldsymbol{\sigma})\,\boldsymbol{\sigma}
}.
\label{eq:dissipative_flow}
\end{equation}

\section{Density flow equation as a gradient flow} \label{app::gradient}
The differential of the time-dependent informational potential $U$ along a traceless direction $\boldsymbol{X}$ is
\begin{equation}
dU_{\boldsymbol{\sigma}}[\boldsymbol{X}]
=
-2\,\operatorname{Tr}(\boldsymbol{Q}\boldsymbol{X}).
\end{equation}
On the subspace of traceless directions, the Riemannian gradient $\nabla_{g} U$ is defined by
\begin{equation}
\boldsymbol{g}_{\boldsymbol{\sigma}}(\nabla_{g} U,\boldsymbol{X})
=
dU_{\boldsymbol{\sigma}}[\boldsymbol{X}],
\quad \forall \, \boldsymbol{X} \quad \text{with} \quad
\operatorname{Tr}\boldsymbol{X}=0.
\end{equation}
This is satisfied by
\begin{equation}
\boldsymbol{K}_{\boldsymbol{\sigma}}(\nabla_{g} U)
=
-2\big(
\boldsymbol{Q}
-
\operatorname{Tr}(\boldsymbol{Q}\boldsymbol{\sigma})\boldsymbol{I}
\big),
\end{equation}
which yields
\begin{equation}
\nabla_{g} U
=
-2\{\boldsymbol{Q},\boldsymbol{\sigma}\}
+
4\,\operatorname{Tr}(\boldsymbol{Q}\boldsymbol{\sigma})\boldsymbol{\sigma}.
\end{equation}
Therefore, the flow \eqref{eq:dissipative_flow} can be written as
\begin{equation}
\boxed{
\dot{\boldsymbol{\sigma}}
=
-\frac12\,\nabla_{g} U(\boldsymbol{\sigma}),
}
\label{eq:gradient_flow}
\end{equation}
showing that the density operator follows a gradient flow with respect to the Bures metric with a time-dependent potential $U$. 

\end{appendices}


\end{document}